\documentstyle[12pt]{article}

\font\SC=cmcsc10 scaled 1440

\def\rdots{\mathinner{\mkern1mu\raise1pt\vbox{\kern1pt\hbox{.}}\mkern2mu
   \raise4pt\hbox{.}\mkern2mu\raise7pt\hbox{.}\mkern1mu}}
\newcommand{\Z}{{\rm Z\kern-.35em Z}}
\newcommand{\bP}{{\rm I\kern-.15em P}}
\newcommand{\Q}{\kern.3em\rule{.07em}{.65em}\kern-.3em{\rm Q}}
\newcommand{\R}{{\rm I\kern-.15em R}}
\newcommand{\h}{{\rm I\kern-.15em H}}
\newcommand{\C}{\kern.3em\rule{.07em}{.55em}\kern-.3em{\rm C}}
\newcommand{\T}{{\rm T\kern-.35em T}}
\newcommand{\D}{{\kern-.5em /}}

\begin{document}

\openup 1.5\jot

\centerline{A Gauge-Invariant Regularization of the}

\centerline{Weyl Determinant Using Wavelets $^*$}

\vspace{1.0in}
\centerline{Paul Federbush}
\centerline{Department of Mathematics}
\centerline{University of Michigan}
\centerline{Ann Arbor, MI 48109-1003}
\centerline{(pfed@math.lsa.umich.edu)}
\vspace{4.0in}

$^*$ This work was supported in part by the National Science Foundation under
Grant No. PHY-92-04824.

\vfill\eject

\centerline{ABSTRACT}

\indent
In line with a previous paper, a gauge-invariant regularization is developed
for
the Weyl determinant of a Euclidean gauged chiral fermion.  We restrict
ourselves to gauge configurations with the $A$ field going to zero at infinity
in
Euclidean space; and thus restrict gauge transformations to those with $U$
the identity at infinity.  For each finite cutoff one gets a strictly
gauge-invariant
expression for the Weyl determinant.  Full Euclidean invariance is only to be
sought in the limit of removing the cutoff.  We expect the limit to be
Euclidean invariant, but this has not yet been proved.   One need not enforce
the no-anomaly condition
on the representation of the gauge group!  We leave to future research relating
the present results to conventional physics wisdom.

\vfill\eject

{\SC INTRODUCTION}

\indent

Following the development of a gauge-invariant regularization for a gauged
scalar boson, and a gauged dirac fermion in [1], we continue in a similar vein
with the treatment of a chiral gauged fermion by study of the Weyl determinant.
The most surprising aspect of the study is that we are not forced to impose a
no-anomaly condition on the group representation (at least for the results
obtained so far).

The traditional treatment of the Weyl determinant is by Leutwyler in [2].
Further
research will be required to relate our results to the usual textbook
statements.
But it seems certain that at the very least new insights into the role of
anomalies
will be uncovered.

\bigskip
\indent
{\SC AN ASIDE}

It perhaps should be noted that there are ``cheap'' ways to obtain
gauge-invariant regularizations.  For example, one could define the determinant
as to be calculated in the radial axial gauge about the origin.  This is
tautologically a gauge invariant definition.  (It is blatantly not Euclidean
invariant.)  Then one can employ momentum cutoffs.  But this procedure is very
far from having the {\it localization} property described in Remark 3) at the
end of the paper.  There is no reason to expect renormalization to be
implementable using {\it local} counterterms.

\bigskip
{\SC BASIC FACTS ABOUT THE WEYL OPERATOR}

\indent
In four dimensional Euclidean space we study the Weyl operators

\begin{equation}
W_\pm = \sum^3_{j=1} \sigma_j(\partial_j + A_j) \pm i(\partial_0 + A_0)
\end{equation}
with $\sigma_j$ the Pauli matrices.  The $A_\mu$ are anti-hermitian, that is
$iA_\mu$ is hermitian.  With $*$ representing conjugate-transpose one has
\begin{equation}
W^*_\pm = -W_\mp \ .
\end{equation}
We first briefly consider the special case of a real representation
of the gauge group.
That is, the representation is equivalent to its conjugate.  For simplicity we
write the following two equations, (3) and (4), in the special case when
$A_\mu$ is real.  Using $c$ to denote conjugation we then get
\begin{equation}
\sigma_2 \ W^c_\pm \ \sigma_2 = - W_\pm \ .
\end{equation}
{}From (2) and (3) we see that for a real representation one has
\begin{equation}
 \sigma_2 \ W^T_\pm \ \sigma_2 =  W_\mp
\end{equation}
with $T$ indicating transpose.  For a real representation thus one has
\begin{equation}
\det (W_+) = \det (W_-)
\end{equation}
and one can obtain the Weyl determinants from the square root of the Dirac
operator determinant.  For non-real representations only the magnitude of the
Weyl determinant may be deduced from the Dirac determinant.  Explicitly
\begin{equation}
i\; D \ = \ i
\left(
\begin{array}{cc}
0 & W_+ \\
W_- & 0
\end{array}
\right)
\end{equation}
in a suitable representation of the $\gamma$ matrices.  And it follows
\begin{equation}
\det(i\; D ) = det(W_+) \det(W_-)
\end{equation}
from which the relationship between the Weyl and Dirac determinants.  In what
follows we study $W_+$ abbreviated as $W$.

\bigskip

\indent
{\SC THE WAVELET BASES}

We follow the notation of Chapter 2 of [1].  We let $\psi_\alpha(x)$ be the
wavelet basis constructed by Y. Meyer, orthonormal in the usual inner product.
\begin{equation}
\left< \psi_\alpha, \psi_\beta \right> = \int d^4x \; \psi_\alpha(x) \;
\psi_\beta(x) = \delta_{\alpha, \beta}.
\end{equation}
The $\psi_\alpha$ carry suppressed 2-spinor and group indices.  We let
$\psi^1_\alpha$, (a non-orthonormal basis) be defined by
\begin{equation}
\psi^1_\alpha = {1 \over W_0} \; \psi_\alpha
\end{equation}
with $W_0$ the operator
\begin{equation}
W_0 = \sum^3_{j=1} \sigma_j \partial_j + i \partial_0 \ .
\end{equation}
The Weyl determinant is then first viewed as the determinant of the infinite
discrete matrix $M_{\alpha,\beta}$:
\begin{eqnarray}
\det(W) &=& \det(M_{\alpha,\beta}) \\
M_{\alpha,\beta} &=& \left< \psi_\alpha, \; W \; \psi^1_\beta \right> .
\end{eqnarray}

\bigskip

{\SC THE CHANGE OF BASIS}

\indent
We follow [1] and now change the bases used in (12), defining
\begin{equation}
\phi_\alpha(x) = u(x,\gamma_\alpha) \psi_\alpha(x)
\end{equation}
and
\begin{equation}
\phi^1_\alpha(x) = u(x, \gamma_\alpha)\psi^1_\alpha(x)
\end{equation}
These definitions mimic equation (4.4) in [1] using (3.8) of [1] to define
$u(x,\gamma_\alpha)$, and recalling from Chapter 2 of [1] that $\gamma_\alpha$
is the ``center'' of $\psi_\alpha$.  Each wavelet has been put in the radial
axial gauge about its center by the gauge transformations
in (13) and (14).  It must be noted that the gauge transformation is different
from wavelet to wavelet, since different wavelets (may) have different centers.
 Changing bases in (12) we get now for the Weyl determinant:
\begin{equation}
\det(W) = \det(M_{\alpha,\beta}) \ = \ \det(N_{\alpha,\beta)} \big/
\Big(\det(A_{\alpha,\beta}) \det(B_{\alpha,\beta})\Big)
\end{equation}
with
\begin{eqnarray}
N_{\alpha,\beta} &=& \Big< \phi_\alpha , W \phi^1_\beta \Big> \\
A_{\alpha,\beta} &=& \Big< \phi_\alpha , \psi_\beta \Big> \\
B_{\alpha,\beta} &=& \Big< W^*_0 \psi_\alpha ,  \phi^1_\beta \Big> \ .
\end{eqnarray}
We now seek gauge-invariant representations for the numerator and denominator
in (15), that is gauge-invariant cutoffs for the corresponding infinite
determinants.

\bigskip

{\SC THE NUMERATOR DETERMINANT}

\indent
If we restrict the indices in $N_{\alpha,\beta}$ to any finite set, the
truncated matrix $N^{TR}_{\alpha,\beta}$ has a gauge-invariant determinant.
This is as in the study of $n_{\alpha,\beta}$ in [1], in equations
(4.12)-(4.23) therein.  Actually with our conditions on the potential
$A_\mu(x)$ we need only an ultraviolet cutoff, restricting ourself to wavelets
with length scale, $\ell$, greater than some cutoff, $\ell_0$.

\bigskip

{\SC THE DENOMINATOR DETERMINANTS}

\indent
We proceed to study the denominator in (15).  We first define
\begin{equation}
C_{\alpha,\beta} = \Big< \psi_\alpha, \phi_\beta \Big>
\end{equation}
\begin{equation}
D_{\alpha,\beta} = \Big< \phi_\alpha, \phi_\beta \Big> .
\end{equation}
We write the denominator determinants
\begin{equation}
\det(A) \det(B) = L \cdot R
\end{equation}
with
\begin{equation}
L = \det(A) \det(C) = \det(D)
\end{equation}
and
\begin{equation}
R = \det(B)/ \det(C) = {\det(\left< W^*_0 \psi_\alpha \; , \; \phi^1_\beta
\right>) \over \det(\left< \psi_\alpha \; , \; \phi_\beta \right> )} .
\end{equation}
$L$ is simple to deal with; for any finite truncation of $D$, $\det(D)$ is
gauge-invariant similar to $B(0,0)$ in (4.28) of [1].  The treatment is as in
(4.38)-(4.43) of [1].  Again we will only need an ultraviolet cutoff.

The study of $R$ is more difficult and more interesting .  One would like to
imitate the development in [1] beginning with equation (4.24).  There one
interpolated between $(-\Delta + m^2)$ and $1$ using operators $(-\Delta +
m^2)^s, \ 0 \le s \le 1$.  It is the lack of a similar suitable interpolation
between $W_0$ and $1$ that makes the chiral situation less elegant.  But we
begin with a preliminary interpolation.

\bigskip

{\SC AN INTERPOLATION STEP}

\indent

We write $R$ as a quotient
\begin{equation}
R = {X \over Y}
\end{equation}
with
\begin{eqnarray}
X &=& \det \left( \big< W^*_0 \psi_\alpha, \phi^1_\beta \Big> \right) / \det
\left( \big<  \psi^{(-1)}_\alpha, \phi ^{(1)}_\beta \Big> \right)  \\
Y &=& \det \left( \big<  \psi_\alpha, \phi _\beta \Big> \right) / \det \left(
\big<  \psi^{(-1)}_\alpha, \phi ^{(1)}_\beta \Big> \right)
\end{eqnarray}
having introduced
\begin{eqnarray}
\psi^{(s)}_\alpha &=& {1 \over (-\Delta)^{s/2}} \ \psi_\alpha \\
\phi^{(s)}_\alpha &=& u(x,\gamma_\alpha) \ \psi^{(s)}_\alpha \ .
\end{eqnarray}
These interpolating functions enable $Y$ to be  treated as in [1], and yield a
manifestly gauge-invariant development for $Y$, with ultraviolet cutoffs
yielding a gauge-invariant regularization.  We have isolated all the new
features and difficulties into $X$.

\bigskip

{\SC THE NEW FEATURE}

\indent
Again we write $X$ as a ratio
\begin{equation}
X = {E \over F}
\end{equation}
with
\begin{eqnarray}
F &=& \det \left( \big< \phi^{(-1)}_\alpha, \psi^{(1)}_\beta \Big> \right)
\det \left( \big<  \psi^{(-1)}_\alpha, \phi ^{(1)}_\beta \Big> \right)  \\
&=& \det \left( \Big< \phi^{(-1)}_\alpha, \phi^{(1)}_\beta \Big> \right)
\end{eqnarray}
and
\begin{eqnarray}
E &=& \det \left( \big< \phi^{(-1)}_\alpha, \psi^{(1)}_\beta \Big> \right)
\det \left( \big<  W^*_0 \psi_\alpha, \phi^1_\beta \Big> \right)  \\
&\sim& \det \left( \Big< \phi^{(-1)}_\alpha, \phi^1_\beta \Big> \right)
\end{eqnarray}
$F$ is as simple to treat as $\det(D)$ before, immediately of the form
developed in [1].  The proportionality in (33) indicates a numerical factor
independent of the gauge field.

Collecting the expression we have obtained so far for the Weyl determinant
\begin{equation}
\det(W) \ \sim \ {\det(N) \cdot Y \cdot  F \over \det(D) \cdot E} \ \ .
\end{equation}

Here $\det(N), \ \det(D), \ F$, and $Y$ are all in what may be called a
``standard form''.  They are all types of expressions met in [1].  Each is a
gauge-invariant expression that may be ultraviolet cutoff by eliminating
wavelets below some length scale and yielding a gauge-invariant cutoff
regularization.  (Each such truncation is gauge-invariant.)  In these
developments one always is working with matrices that are the identity if the
gauge field is zero.  One finds the gauge field contributions as traces of
closed line integrals of the gauge field, with possible $F$ field inserts,
manifestly gauge-invariant expressions.

$E$ is more complicated than the matrices treated in [1].  Any truncation of it
is gauge- invariant, but it does not become the identity if the gauge field is
zero.  We let $E^{TR}_0$ be the corresponding matrix with the gauge field set
zero.  We also set
\begin{equation}
V = E^{TR} - E^{TR}_0 .
\end{equation}

We then have
\begin{eqnarray}
\det(E^{TR}) &=& \det(E^{TR}_0 + V) \\
&=&  \det(E^{TR}_0)  \det \left( 1 + (E^{TR}_0)^{-1} \; V \right) \\
&\sim&  \det \left( 1 + (E^{TR}_0)^{-1} \; V \right)
\end{eqnarray}
This is a good expression from which to compute contributions of $E$ in
perturbation theory.  But, (38) is not as friendly an expression to get
estimates from as the ``standard forms'' met in [1].  (There are alternate ways
to treat $E$ other than the development in (38).)  {\bf If} anomalies rear
their ugly head, they will arise {\bf only} from treachery concealed in $E$.

\bigskip
{\SC THE MATRIX  $(E^{TR}_0)^{-1}$}

\indent
We abbreviate $(E^{TR}_0)$ as $Z$.  We see that
\begin{equation}
Z_{\alpha,\beta} = \bigg< \psi_\alpha \; , \  {(-\Delta)^{1/2} \over W_0} \
\psi_\beta  \bigg>
\end{equation}
We proceed to find a convenient expression for the inverse of $Z$.  We define
\begin{equation}
Q_{\alpha,\beta} = \left< \psi_\alpha, \ {W_0 \over (-\Delta)^{1/2}} \
\psi_\beta \right> .
\end{equation}
$Q_{\alpha ,\beta}$ is taken with the same truncation as $Z_{\alpha, \beta}$.
(They are restricted to the same subset of wavelets.)  If there were no
truncation, $Z$ and $Q$ would be inverses (as well as conjugate transposes) of
each other.  It is natural to then write
\begin{equation}
Z^{-1} \ = \ (1 + e)Q
\end{equation}
where $e$ is a ``small'' matrix, zero if no truncation.  We then have
\begin{eqnarray}
Z^{-1}Z &=& I \\
(1+e)QZ &=& I \\
e &=& (I-QZ)(QZ)^{-1} .
\end{eqnarray}

If we take a ``sharp'' ultraviolet cutoff, keeping all wavelets with length
scales, $\ell $, such that $\ell \ge \ell_0 = {1 \over 2^{r_0}}$,  and
discarding wavelets with length scales $\ell < \ell_0$, we find the following
properties of our matrices:

1)  $(I-QZ)_{\alpha,\beta}$ is zero unless both $\alpha$ and $\beta$ are at
level $\ell_0$.  This fact depends on the property that Y. Meyer wavelets have
of having no overlap in momentum space between wavelets differing by more than
one level, and the diagonality of $(-\Delta)^{1/2} / W_0$ in momentum space.

2)  $(QZ)_{\alpha, \beta}$ is the identity for $\alpha, \beta$ at length scales
$\ell > \ell_0$, has zero coupling between levels $\ell > \ell_0$ and level
$\ell_0$.

It follows that $Z^{-1}$ couples wavelets with length scales differing at most
by one level.  To obtain good estimates we also want that
\begin{equation}
|Z^{-1}_{\alpha, \beta}|  \le  c_n {\ell^n_\alpha \over |\gamma_\alpha -
\gamma_\beta|^n}
\end{equation}
for all $n > 0$ and some set of $c_n$.  That is, we want matrix elements to
fall off faster than any power of the distance between the centers of the
wavelets, as measured in the length scale of the wavelets.  This estimate is
certainly true except possibly when either $\alpha$ or $\beta$ are at the
bottom level, $\ell_0$ (since $Q$ satisfies the estimate).  To ensure this
estimate we modify the truncation of $E$ so that
\begin{description}
\item [1)] we keep all wavelets with $ \ell > \ell_0$
\item [2)] discard all wavelets with  $\ell < \ell_0$
\item [3)] at level $\ell_0$ we keep half the wavelets, the black squares of a
checkerboard pattern.
\end{description}
We leave to a later publication showing that this ensures estimate (45).  (It
is not necessary for our other truncations to share this modification from a
``sharp'' cutoff.)

\bigskip
{\SC FINAL REMARKS}

\indent
We restrict our observations to the perturbative regime.
\begin{equation}
\ell n \; \left(\det(W) \right) \ = \ T_1 + T_2 + \cdots
\end{equation}
where $T_n$ is a homogeneous polynomial of degree $n$ in the $A_\mu$ field.

1)  For $n \ge 5$ the gauge-invariant cutoff (regularized) $T_n$ converge as
the cutoff is removed.  This is easy.  We will want to prove that the limit for
$T_n$ agrees with any other calculation of these terms, that do not require
renormalization.

2)  For $n \le 4$ we have gauge-invariant cutoff (regularized) expressions for
the $T_n$.  We want to compute the limits, subtracting gauge-invariant
counterterms  if necessary, and prove the limits are Euclidean invariant.
These may be difficult computations.  {\bf If} there are any difficulties with
anomalies, it will be here.

3)  We wish finally to emphasize a feature of our regularizations, following
from properties of wavelets and the constructions employed.  If one looks at
two different cutoffs $\ell_0$ and $\ell '_0$ $< \ell_0$, then the cutoff
expressions for $T_n$,  $T_n(\ell_0)$ and  $T_n(\ell '_0)$, will differ by
terms localized on a length scale $\ell_0$.  Connected diagrams for the
difference have kernels rapidly going to zero when vertices separate measured
on length scale $\ell_0$ analogous to equation (45)).  This was one goal of our
constructions.

\vfill\eject

\centerline{{\SC REFERENCES}}

\begin{description}
\item [[1]] P. Federbush, "A New Formulation and Regularization of Gauge
Theories Using a Non-Linear Wavelet Expansion", preprint (hep-ph/9505368).

\item[[2]] L. Leutwyler, "On the Determinant of the Weyl Operator", in
{\it Quantum Field theory and Quantum Statistics}, I.A. Batalin, C.J. Isham and
G.A. Vilkovisky (eds.), A. Hilger , 1987.

\end{description}

\end{document}